\magnification=\magstep1
\tolerance=500
\rightline{TAUP 2541-99}
\rightline{17 December, 1998}
\vskip 2cm
\centerline{\bf Chaos and Maps in Relativistic Dynamical Systems}
\smallskip
\centerline{L.P. Horwitz$^{a,b}$ and Y. Ashkenazy$^b$}
\smallskip
\centerline{${}^a$Raymond and Beverly Sackler Faculty of Exact Sciences}
\centerline{School of Physics, Tel Aviv University, Ramat Aviv 69978, Israel}
\centerline{${}^b$Department of Physics}
\centerline{Bar Ilan University, Ramat Gan 52900, Israel}
\bigskip
\noindent
{\it Abstract:\/} The basic work of Zaslavskii {\it et al} showed that
the classical non-relativistic electromagnetically kicked oscillator
can be cast into the form of an iterative map on the phase space; the
resulting evolution contains a stochastic flow to unbounded
energy. Subsequent studies have formulated the problem in terms of a
relativistic charged particle in interaction with the electromagnetic
field.
We review the structure of the covariant Lorentz force used to study
this problem. We show that the Lorentz force equation can be derived
as well from the manifestly covariant mechanics of Stueckelberg in the
presence of a standard Maxwell field, establishing a connection
between these equations and mass shell constraints. We argue that these
relativistic generalizations of the problem are intrinsically
inaccurate due to an inconsistency in the structure of the
relativistic Lorentz force, and show that a reformulation of the
relativistic problem,  permitting variations (classically) in both
the particle mass and the effective ``mass'' of the interacting
 electromagnetic field, provides a consistent system of classical
equations for describing such processes.
\vfill
\break
{\bf 1. INTRODUCTION}
\smallskip
\par Zaslavskii {\it et al} [1] have studied the behavior
of particles
in the field of a wave packet of an electric field in the presence of a
static magnetic field.
For a broad wave packet with sufficiently uniform spectrum,
the problem can be stated in terms of an electrically kicked harmonic
oscillator. They find that for rational ratios between the frequency
of the kicking field and the Larmor frequency associated with the
magnetic field, the phase space of the system is covered by a mesh of finite
thickness; inside the filaments of the mesh, the dynamics of the particle is
stochastic and
outside (in the cells of stability), the dynamics is regular. This structure is
called a stochastic web. It was found that this pattern covers the entire
phase plane, permitting the particle to diffuse arbitrarily far into the
region of high energies (a process analogous to Arnol'd
diffusion [2]).
\par Since the stochastic web leads to unbounded energies, several
authors have
considered the corresponding relativistic problem. Longcope and Sudan [3]
 studied this system (in effectively
$1+1/2$ dimensions) and found that for initial conditions close to the
origin of the phase space there is a stochastic web, which is bounded in
energy, of a form quite similar, in the neighborhood of the origin, to the
non-relativistic case treated by
Zaslavskii {\it et al} [1]. Karimabadi and Angelopoulos [4]
studied the case of an obliquely propagating wave, and showed that under
certain conditions, particles can be accelerated to unlimited energy through
an Arnol'd diffusion in two dimensions.
\par The equations used by Longcope and Sudan[3] and Karimabadi and
 Angelopoulos[4] are derived from the well-known covariant Lorentz
 force
$$ f^\mu = m {d^2 x^\mu \over ds^2} = {e \over c} F^\mu\,_\nu {dx^\nu
\over ds}, \eqno(1.1)$$
where $ds$ is usually taken to be the ``proper time'' of the particle.
Multiplying both side by $dx_\mu / ds$ and summing over $\mu$ (we use
the Einstein summation convention that adjacent indices are summed
unless otherwise indicated, and the metric is taken to be
$(-,\,+,\,+,\,+)$ for the indices $(0,\,1,\,2,\, 3)$, distinguishing 
 upper and lower indices), one obtains
$$ {dx_\mu \over ds} {d^2x^\mu \over ds^2} = {1 \over 2} {d \over ds}
\bigl( {dx_\mu \over ds}{dx^\mu \over ds} \bigr) = 0; \eqno(1.2)$$
taking the usual value for the constant (in $s$), we have that
$$ {dx_\mu \over ds}{dx^\mu \over ds} = -c^2.  \eqno(1.3)$$
This result provides a consistent identification of  the parameter $s$
on the particle trajectory (world-line) as the ``proper time'':
$$ \eqalign{ds^2 &= -{1 \over c^2} dx_\mu dx^\mu \cr
&= dt^2 - {1 \over c^2} d{\bf x}^2 \cr
&= dt^2 \bigl( 1 - {v^2 \over c^2} \bigr)\cr} \eqno(1.4) $$
so that
$$ dt = { ds \over \sqrt{1 - {v^2 \over c^2}}} \equiv \gamma
ds, \eqno(1.5)$$
the Lorentz transformation of the time interval for a particle at rest
to the interval observed in a moving frame.
This formula has been used almost universally in calculations of the
dynamics of relativistic charged particles [6,7]. The Lorentz
trtansformation, however, applies only to inertial frames.  Phenomena
occurring in two inertial frames in relative motion are, according to
the theory of special relativity, related by a Lorentz
transformation.  An accelerating frame, as pointed out by Landau and
Lifshitz[6], induces a more complicated form of metric than the flat
space $(-\,,+\,,+\,,+)$.  Mashoon [11] has emphasized that the use of
a sequence of instantaneous inertial frames, as has also often been done, is
not equivalent to an accelerating frame.  He cites the example for
which a  charged particle at rest
in an inertial frame does not radiate, while a similar particle at
rest in an accelerating frame does. As another example, consider
again the first of  $(1.4)$. If we transform to the
inertial frame of a particle with constant acceleration along the $x$
 direction,
$$x' = x +{1 \over 2}a t^2, $$
then $(1.4)$ becomes (as in the discussion of rotating frames in [8])
$$ ds^2 = ( 1 - {1 \over c^2}a^2t^2) dt^2 - {2 \over c^2}atdx'dt
-{1 \over c^2}(dx'^2 +dy^2).$$
In the frame in which $dx'=dy=dz=0$, $dt$ is the interval of proper
time, and it is not equal to $ds$.  For short times, or small
acceleration, the effect is small.  We shall discuss this problem
further in Section 3.
\par   Continuing for now in the standard framework, Eq. $(1.3)$ effectively
eliminates one of the equations of $(1.1)$.  We may write
$$ \eqalign{{d^2 {\bf x} \over ds^2} &= \bigl({dt \over ds} \bigr) {d \over dt}
\bigl( {dt \over ds}{d{\bf x} \over dt} \bigr) \cr
&= {e \over m} {dt \over ds} ({\bf E} + {1 \over c} {\bf v} \times
{\bf H}).\cr} $$
Cancelling $\gamma = dt/ds$ from both sides, one obtains
$$ {d \over dt} (\gamma {\bf v}) = {e \over m} ({\bf E} + { 1\over c}
{\bf v} \times {\bf H}), \eqno(1.6)$$
the starting point for the analysis of Longcope and Sudan [3] and
Karamabadi and Angelopoulis [4].  A discrete map can be constructed
from $(1.6)$ just as was done for the nonrelativistic equations of
Zaslavskii {\it et el} [1].  As we have remarked above, the stochastic web
is found at low energies; it deteriorates at high energies due to the
$\gamma$ factor.
\par The time component of $(1.1)$ is
$$ c { d^2t \over ds^2} = {e \over mc} {\bf E}\cdot {\bf v} {dt \over
ds} \eqno(1.7)$$
or
$$ {d\gamma \over dt} = {e \over mc^2}  {\bf E}\cdot {\bf v}.
\eqno(1.8)$$
Landau and Lifshitz [6] comment that this is a reasonable result,
since the ``energy'' of the particle is $\gamma mc^2$, and
 $ e{\bf E}\cdot {\bf v}$ is the work done on the particle by the
field. It is important for what we have to say in the following that
Eq. $(1.7)$ is not interpretable in terms of the geometry of Lorentz
transformations. The second
derivative corresponds to an {\it acceleration} of the observed
time variable relative to the ``proper time''; the Lorentz
transformaton affects only the first derivative, as in $(1.4)$. We
understand this equation as an indication that the observed time
emerges as a dynamical variable.  Mendon\c ca and Oliveira e Silva [8]
have studied the relativistic kicked oscillator by introducing a
``super Hamiltonian'', resulting in a symplectic mechanics of
Hamiltonian form, which recognizes that the variables $t$ and $E$ are
dynamical variables of the same type as ${\bf x}$ and ${\bf p}$.  This
manifestly covariant formulation is equivalent to that of
Stueckelberg [9] and Horwitz and Piron [10], which we shall discuss
in the next section.
\par We have computed solutions to the Lorentz force  equation for the
case of the kicked oscillator (see fig. 1), using methods slightly
different from that of Longcope and Sudan [3] and Karimabadi and
Angelopoulis [4].  At low velocities, the
stochastic web found by Zaslovskii {\it et al} [1] occurs; the system diffuses
in the stochastic region to unbounded energy, as found by Karimabadi and
Angelopoulis [4]. The velocity of the particle is light
speed limited by the dynamical equations, in particular, by the suppression
of the action of the electric field at velocities approaching the velocity
of light [5].
\par The rapid acceleration of the charged particles of the kicked
 oscillator further suggest that radiation can be an important
 correction to the motion. The counterexample of Mashoon [11] was
 based on the phenomenon of radiation.  It has been shown by
Abraham [12], Dirac [13], Rohrlich [6]
 and Sokolov and Ternov [14] that the relativistic  Lorentz force
 equations in the presence of radiation reaction
 is given by the Lorentz-Abraham equation
$$ m {\ddot x}^\mu =
 {e \over c} F^{\mu \nu}{\dot x}_\nu + {2 \over 3} {r_0
\over c} m ({d \over ds}{\ddot x}^\mu - {1 \over c^2} {\dot x}^\mu {\ddot
x}_\nu {\ddot x}^\nu ),\eqno(1.9) $$
where  $r_0 = e^2 /mc^2$, the classical electron radius, and the dots
refer here, as in $(1.1)$, to derivatives with respect to $s$.  Note
that from the identity $(1.3)$, it follows (by differentiation with
respect to $s$) that
$${ \dot x}_\mu {\ddot x}^\mu = 0, \qquad {\dot x}_\mu
{d \over ds}{\ddot x}^\mu + {\ddot x}_\mu {\ddot x}^\mu = 0,
\eqno(1.10)$$
and hence $(1.9)$ can be written as
$$ m {\ddot x}^\mu =
 {e \over c} F^{\mu \nu}{\dot x}_\nu + {2 \over 3} {r_0
\over c} m {d \over ds}{\ddot x}^\nu(\delta^\mu_\nu + {1 \over c^2} {\dot
x}^\mu {\dot x}_\nu).\eqno(1.11) $$
The last factor on the right is a projection orthogonal to ${\dot
x}^\mu$ (if ${\dot x}^\mu{\dot x}_\mu = -c^2$), and therefore $(1.11)$
is consistent with conservation of ${\dot x}^\mu {\dot
x}_\mu$. Sokolov and Ternov [4] state that this conservation law
follows automatically from $(1.9)$, but it is apparently only
consistent. Radiation reaction therefore also implies that the
connection between proper time and the Lorentz invariant interval may
be subject to question.
\par We have calculated the motion of the  kicked oscillator using the
form $(1.9)$ of the Lorentz force, corrected for radiaton reaction,
 undoubtedly a good approximation  under certain conditions,
 and will
 report on this in another paper in this volume [15].
\bigskip
{\bf 2. THE STUECKELBERG FORMULATION}
\smallskip
\par  As we have remarked above, Mendon\c ca and Oliveira e Silva [8]
have used a ``super Hamiltonian'' formulation to control the
covariance of the electromagnetically kicked oscillator.  Their
formulation of the problem is equivalent to the theory of
Stueckelberg [9] and Horwitz and Piron [10]; we shall therefore use
the notation of the latter formulation. We first explain the physical
basis of this theory, and then derive sytematically the covariant
Lorentz force from a model Hamiltonian.
\par The original thought experiment of Einstein [16] discussed the
generation of a sequence of signals in a frame $F$, according to a clock
imbedded in that frame, and their detection by apparatus in a second frame
 $F'$ in uniform motion with respect to the first. The time of arrival
of the signals in $F'$ must be recorded with a clock of the same
construction, or there would be no basis for comparison of the
intervals between signals sent and those received.  It is essential to
understand that the clocks in both $F$ and $F'$ run at the same rate.
The relation of the interval $\Delta \tau$ between pulses emitted in $F$
and the interval between signals $\Delta \tau'$ received in $F'$,
according to the (equivalent) clock in $F'$ is, from the special theory
of relativity, given by
$$\Delta \tau' = {\Delta \tau \over \sqrt{ 1 - {v^2 \over c^2}}}.
\eqno(2.1)$$
This time interval, measured on a ``standard'' time scale established
by these equivalent clocks, is identified to the interval $\Delta t'$,
the time interval between signals in the first frame, observed in the
second, and called simply the {\it time} by Einstein.  One sees that
this Einstein time is subject to distortion due to motion.  In general
relativity, it is subjuct to distortion due to the gravitational field
 as well,
and in this case the distortion is called the gravitational
red-shift. We see that there are essentially {\it two types of time}, one
corresponding to the time intervals at which signals are emitted, and
the second, according to the time intervals for which they are detected.
The first type, associated with signals that are pre-programmed, is
not a dynamical variable, but a given sequence (as for the Newtonian
time), and the second, associated with the time at which signals are
observed (the Einstein time), is to be understood as a dynamical
variable both in classical and quantum theories [17].
\par Stueckelberg [9] noted that for a free particle, the signals emitted
at regular intervals would be recorded at regular intervals in a
laboratory, since the free particle would be in motion with respect to
the laboratory with the same relation as between $F$ and $F'$; the
motion would then be recorded as a straight line (within the light
cone) on a spacetime diagram.  In the presence of forces, however,
this line could curve.  A sufficient deviation from the straight line 
 could make it 
begin to go backward in time, and then the coordinate $t$ wuld no
longer be adequate to parametrize the motion.  He therefore introduced
an invariant parameter $\tau$ along the curve, so that there would be
a one-to-one corrrespondence between this parameter and the spacetime
coordinates. He proposed a Hamiltonian for a free particle of the form
(the parameter $M$ provides a dimensional scale, for example, in
$(2.5)$; it may also be considered as the Galilean target mass for the
{\it variable} $(1/c)\sqrt{E^2 - c^2 {\bf p}^2}$)
$$ K= {p^\mu p_\mu \over 2M}  \eqno(2.2)$$
for which the Hamilton equations (generalized) give
$$ {dx^\mu \over d\tau} = {\partial K \over \partial p_\mu} = {p^\mu
\over M}.\eqno(2.3)$$
It is clear that such a theory is intrinsically ``off-shell'';
the variables ${\bf p}$ and $p^0 = E/c$ are independent, as are the
observables $\bf x$ and $t$, so that the phase space is eight-dimensional.
Dividing the equation for the space indices by the equation for the
time index, one obtains
$$ {\bf v} = {d{\bf x} \over dt} = c^2 {{\bf p} \over E},\eqno(2.4)$$
precisely the Einstein formula for velocity.  Furthermore, for the
time component,
$$ {dt \over d\tau} = {E \over Mc^2}; \eqno(2.5)$$
in case the particle is ``on-shell'', so that $Mc^2 = \sqrt{E^2 -
c^2 {\bf p}^2}$, $(2.5)$ reads
$$ {dt \over d\tau} = {1 \over \sqrt{1 - {{\bf v}^2 \over c^2}}},$$
coinciding with $(2.1)$.
\par Stueckelberg [9] then considered adding a potential term $V(x)$,
to treat one-body mechanics, and the gauge substitution $p^\mu
\rightarrow p^\mu -eA^\mu(x)$ for the treatment of problems with
electromagnetic interaction.  He proposed a quantum theory, for which
the Hamiltonian generates a Schr\"odinger type evolution
$$ i\hbar {\partial\over \partial \tau} \psi_\tau (x) = K
\psi_\tau(x). \eqno(2.6)$$
\par Horwitz and Piron [10] generalized the Stueckelberg theory for
application to many body problems.  They assumed that the standard
clocks constitute a universal time, as for the Robertson-Walker time
(the Hubble time) of general relativity [18],
so that separate subsystems are
correlated in this time. In this framework, it became possible to
solve, for example,
the two body problem in both classical [10] and quantum theory [19].
\par The equations $(1.1)$ are not generally derived rigorously
from a well-defined Lagrangian or Hamiltonian.  They result from a
relativistic generalization of the nonrelativistic Lorentz force
(which is derivable from a nonrelativistic Hamiltonian).  In the
following, we shall derive these equations rigorously from the
Stueckelberg theory, to emphasize more strongly the nature of the
problem we have discussed above, and to clarify some important points.
 \par The Hamiltonianian form for a particle with electromagnetic
interaction proposed by Stueckelberg [9] is
$$ K = {(p^\mu -{e \over c}A^\mu(x) )
(p_\mu - {e \over c}A_\mu(x)) \over 2M}. \eqno(2.7)$$
The equation of motion for $x^\mu$ is (we use the upper dot from now
on to denote differentiation by $\tau$, the universal invariant time)
$$ {\dot x}^\mu = {\partial K \over \partial p_\mu} = {(p^\mu
-{e \over c}A^\mu(x) ) \over M} \eqno(2.8)$$
and we see that then
$$ {dx^\mu \over d\tau}{dx_\mu \over d\tau}=   -c^2 \bigl( {ds \over
d\tau} \bigr)^2 = {(p^\mu -{e \over c}A^\mu(x) )
(p_\mu - {e \over c}A_\mu(x)) \over  M}, \eqno(2.9)$$
a quantity proportional to $K$, and therefore strictly conserved.  In
fact, this quantity is the gauge invariant mass-squared:
 $$ (p^\mu -{e \over c}A^\mu (x) )(p_\mu - {e \over c}A_\mu (x))=
-m^2c^2, \eqno(2.10)$$
where we define $m$ as the dynamical mass, a constant of the motion.
 It then follows that
$$ c^2 \bigl( {ds \over d\tau}\bigr)^2 = c^2 \bigl( {dt \over d\tau}\bigr)^2
-  \bigl( {d{\bf x} \over d\tau}\bigr)^2 = {m^2 c^2 \over M^2}
\eqno(2.11)$$
and, extracting a factor of $(dt/d\tau)$,
$$ \bigl( {dt \over d\tau}\bigr)^2 = {m^2/M^2 \over {1 - {v^2 \over
c^2} }}. \eqno(2.12)$$
Up to a constant factor, the Stueckelberg theory therefore
maintains the identity $(1.3)$.
\par We now derive the Lorentz force from the Hamilton equation (this
derivation has also  been carried out by C. Piron [20]).  The Hamilton
equations for energy momentum are
$$\eqalign{ {dp^\mu \over d\tau} &= - {\partial K \over \partial
x_\mu} = {(p^\nu -{e \over c} A^\nu) \over M } \bigl( {e \over c}
{\partial A_\nu \over \partial x_\mu} \bigr) \cr
&= {e \over c} {dx^\nu \over d\tau} {\partial A_\nu \over \partial
x_\mu} . \cr } \eqno(2.13)$$
Since $p^\mu = M {dx^\mu \over d\tau} + {e\over c} A^\mu$, the
left hand side is ($A^\mu$ is evaluated on the particle world line
$x^\nu(\tau)$)
$$ {dp^\mu \over d\tau } = M {d^2x^\mu \over d\tau^2} + {e \over c}
{\partial A^\mu \over \partial x^\nu} {dx^\nu \over
d\tau},\eqno(2.14)$$
and hence
$$M {d^2 x^\mu \over d\tau^2}= {e \over c} \bigl({\partial A_\nu \over
\partial x_\mu} - {\partial A^\mu \over \partial x^\nu} \bigr) {dx^\nu
\over d\tau},$$
or
$$M {d^2 x^\mu \over d\tau^2} = {e \over c} F^\mu_\nu {dx^\nu \over
d\tau}, \eqno(2.15)$$
where $(\partial^\mu \equiv \partial/\partial x_\mu)$
$$ F^{\mu \nu} = \partial^\mu A^\nu - \partial^\nu
A^\mu. \eqno(2.16)$$
The form of $(2.15)$ is identical to that of $(1.4)$, but the
temporal derivative is not with respect to the variable $s$, the
Minkowski distance along the particle trajectory, but with respect to
the universal evolution parameter $\tau$.
 \par One might argue that these should be equal, or at least
proportional by a constant, since the proper time is equal to the time
which may be read on a clock on the particle in its rest frame.  For
an accelerating particle, however, {\it one cannot transform by a
Lorentz transformation, other
than instantaneously, to the particle rest frame}. It appears,
therefore, that the formula $(2.15)$  could have a more reliable
interpretation.  The parameter of evolution $\tau$ does not require a
Lorentz transformation to achieve its meaning.
\par Since $m^2$ is absolutely conserved by the
Hamiltonian model $(2.7)$, however, we have the constant relation
$$ ds = {m\over M} d\tau, \eqno(2.17)$$
assuming the positive root (as we shall also do for the root of
$(2.12)$; we do not wish to discuss the antiparticle solutions here).
Eq. $(2.15)$ can therefore be written exactly as $(1.1)$.
\par We see that the Stueckelberg formulation in terms of an absolute
time does not avoid the serious problem of consistency that we have
pointed out before.  It is clear that the difficulty is associated
with the fact that the Stueckelberg Hamiltonian, as we have written
it, preserves the mass-shell, and we therefore understand the identity
$(1.3)$ as a {\it mass-shell} relation.
\par Returning to the Stueckelberg-Schr\"odinger equation $(2.6)$, we
see that the gauge invariant replacement    $p^\mu
\rightarrow p^\mu -{e \over c}A^\mu(x)$ is not adequate.  The additional
derivative on the left hand side of $(2.6)$ must also be replaced by a
gauge covariant term. The possibility of $\tau$ dependence in the
gauge transformation implies that the gauge fields themselves may
depend on $\tau$. The gauge covariant equation should then be [21]
$$ i {\partial \over \partial \tau} \psi_\tau(x) =\bigl\{ {1 \over 2M} (p_\mu
-{e_0 \over c} a_\mu)(p^\mu -{e_0 \over c}a^\mu) - {e_0\over c}
a_5\bigr\}
 \psi_\tau(x), \eqno(2.18)$$,
where the fields $a_\alpha,\, \alpha = (0,\,1,\,2,\,3,\,5)$ with $\partial_5 
\equiv \partial/\partial\tau$, change under the gauge transformation
$\psi \rightarrow \exp{i{e_0 \over c}\Lambda}\psi$ according to 
$a_\alpha \rightarrow a_\alpha - \partial_\alpha \Lambda$.
It follows from this equation, in a way analogous to the
Schr\"odinger non-relativistic theory, that there is a current
$$ j_\tau^\mu = -{i \over 2M} \{\psi_\tau^*(\partial^\mu -i{e_0 \over c}a^\mu)
\psi_\tau - \psi_\tau(\partial^\mu + i{e_0 \over c}a^\mu)\psi_\tau^* \},
\eqno(2.19)$$
which, with
 $$\rho_\tau \equiv j^5_\tau = \vert \psi_\tau(x)\vert ^2, $$
satisfies
$$ \partial_\tau \rho_\tau + \partial_\mu j_\tau^\mu \equiv
\partial_\alpha j^\alpha = 0. \eqno(2.20)$$
We see that for $\rho_\tau \rightarrow 0$ pointwise ($\int
\rho_\tau(x) d^4x = 1$ for any $\tau$),
$$ J^\mu(x) = \int_{-\infty}^\infty j_\tau^\mu(x) d\tau \eqno(2.21)$$
satisfies
$$ \partial_\mu J^\mu(x) = 0, \eqno(2.22)$$
and can be a source for the standard Maxwell fields. Since the field
equations are linear, with source $j^\alpha$, one identifies the 
integral $\int \,d\tau a^\mu(x,\tau)$ (or, alternatively, the $0$-mode)
with the Maxwell potentials
[21].  It then follows that the so-called pre-Maxwell fields
$a^\alpha$ have dimension $L^{-2}$, and that the charge $e_0$ has
dimension $L$.  The Lagrangian density for the fields, quadratic in
the field strengths ($\alpha,\beta = 0,\,1,\,2,\,3,\,5$)
 $$f^{\alpha \beta}= \partial^\alpha a^\beta -
 \partial^\beta a^\alpha,$$
which has dimension $L^{-3}$, must carry a dimensional parameter, say
$\lambda$, and from the field equations $\lambda \partial_\alpha f^\beta_\alpha
= e_0 j^\beta$,
 one sees that the Maxwell
charge is $e = e_0 / \lambda$ [21].  
\par We understand the operator on the right hand side of $(2.18)$ as
the quantum form of a classical evolution function
$$ K = {1 \over 2M} (p_\mu
-{e_0 \over c} a_\mu)(p^\mu -{e_0 \over c}a^\mu) - {e_0 \over c} a_5.
 \eqno(2.23)$$
It follows from the Hamilton equations that
$$ {dx^\mu \over d\tau} = {p^\mu -{e_0 \over c}a^\mu \over M}  \eqno(2.24)$$
and
$$ {dp^\mu \over d\tau} = {e_0 \over c} {dx^\nu \over d\tau} {\partial a_\nu
\over \partial x_\mu} + {e_0 \over c} {\partial a_5 \over \partial x_\mu}.$$
Hence,
$$ M{d^2 x^\mu \over d\tau^2} = {e_0 \over c} {dx^\nu \over d\tau} f_\nu^\mu +
{e_0 \over c} \bigl( {\partial a_5 \over \partial x_\mu} - {\partial
a^\mu
 \over
\partial \tau} \bigr) . \eqno(2.25)$$
If we define $x^5 \equiv \tau$, the last term can be written as
$\partial^\mu a_5 - \partial_5 a^\mu = f^\mu_5$ ,
so that
$$ M {d^2x^\mu \over d\tau^2} = {e_0\over c} {dx^\nu \over d\tau}
f_\nu\,^\mu + {e_0 \over c} f_5^\mu. \eqno(2.26)$$
Note that in this equation, the last term appears in the place of the
radiation correction terms of $(1.9)$. It plays the role of a generalized
electric field.  Furthermore, we see that the
relation $(1.3)$, consistent with
 the standard Maxwell theory, no longer holds as an
identity; the Stueckelberg form of this result $(2.11)$ in the
presence of standard Maxwell fields, where $m^2$ is conserved, is also
not generally valid.  We now have
$$ {d \over d\tau } {1 \over 2}M \bigl( {dx^\mu \over d\tau}{ dx_\mu
\over d\tau}\bigr) = {e_0 \over c}  {dx^\mu \over d\tau} f_{\mu 5},
 \eqno(2.27) $$
and does not vanish.  The right hand side corresponds to mass transfer 
from the field to the particle.
\par As for the method of Longcope and Sudan [3], we may  transform
the derivatives with respect to $\tau$ to derivatives with respect to
$t$ in the equation $(2.26)$  as follows. Defining $\zeta = dt/d\tau$,
it follows from $(2.26)$ that there is an additional term in the analogous
form of the rate of change of $\zeta$ (we use lower case to denote the 
pre-Maxwell field strengths),
$$ {d\zeta \over dt} = {e_0 \over Mc^2} ({\bf e} \cdot {\bf v}) + {e_0
\over \zeta Mc^2} f^0\,_5            \eqno(2.28)$$ 
The space components of $(2.26)$ can be written as
$$\eqalign{ {d^2 x^j \over dt^2} &= {e_0 \over \zeta M} \bigl[ e^j + {1 \over c}
({\bf v} \times {\bf h})^j - {v^j \over c^2} ( {\bf e} \cdot {\bf v} )
\bigr] \cr &+ {e_0 \over Mc\zeta^2} \bigl[ f^j\,_5 - {v_j \over c}
f^0\,_5 \bigr]. \cr } \eqno(2.29)$$
\par To illustrate some of the properties of this system of equations, we treat a simple example in Appendix A.
The effective additonal forces include not only the term associated
with the work done by the field, but additonal terms associated
specifically with the $\tau$-dependence of the fields, and the fifth
(scalar) field $a_5$. Given the fields $f^\alpha\,_\beta$, Eqs.$(2.28)$ and 
$(2.29)$ form a nonlinear coupled system of equations for the particle motion.
\par  For a gauge (generalized Lorentz) in which 
$\partial_\alpha a^\alpha =0$, the field equations [21] 
$$ \partial_\alpha f^{\beta \alpha} = e j^\beta $$
become
$$ - \partial_\alpha \partial^\alpha a^\beta = e j^\beta, $$
where, classically, $j^\beta = {\dot x}^\mu \delta^4(x - x(\tau)),
\, \rho = \delta^4 (x - x(\tau))$, and $x \equiv x^\mu (\tau)$ is the
 world line.  
  The analysis of these
equations is in progress.
\par  It has recently been shown that, with the help of the Green's
functions for the wave equations of the fields in $x^\mu, \tau$, that
the self-reaction derived from the contributions on the right hand
side of $(2.26)$ are precisely of the form of the radiation reaction
terms in the Abraham-Lorentz equations $(1.9)$ in the limit that the
theory is constrained to mass shell, {\it i.e.}, that $(1.3)$ is enforced [22].
The off-shell corrections provided by $(2.26)$ make the system of
equations consistent, and should therefore provide a basis for
computing problems involving the interaction of radiation with
relativistic particles in a consistent way.
\bigskip  
{\bf CONCLUSIONS}
\smallskip
\par We have shown that the standard relativistic Lorentz force equations are
not consistent since they imply the mass-shell constraint
 ${\dot x}^\mu {\dot x}_\mu= -c^2$, a relation that can be valid only for a
charged particle moving at constant velocity.  The corrections are
generally small for short times or small accelerations, and therefore
calculations made with this Lorentz force are, in many applications,
quite acceptable. However, for very large accelerations ($at$ large compared
 to $c$), they could become inaccurate.
\par A consistent theory may be constructed from a fully gauge
covariant form of the Stueckelberg [9][10] manifestly covariant
dynamics, a theory which introduces a fifth gauge field [21].  The Lorentz
 invariant force equation derived from this theory contains an
additional term which enters in a way analogous to the radiation
reaction term in the Abraham-Lorentz-Dirac equation (the self
reaction force derived from this generalized equation in the mass-shell
limit coincides with the radiation reaction term obtained by quite
different methods for the Abraham-Lorentz-Dirac equation; it
contains contributions from both terms on the right hand side [22]).
\par It appears that the consistency of the {\it classical} equations
governing the interaction of charged particles with electromagnetic
radiation requires that both the particles and the fields must be
permitted to move ``off-shell'', as in the vertices of quantum field theory.
 \bigskip
\noindent
{\bf Acknowledgements}
\smallskip
\par We thank J. Beckenstein, E. Comay and F. Rohrlich for discussions.
 \bigskip
 \noindent
{\bf Appendix A}
\par The purpose of the following example is to show that in some cases 
the fifth
field ${f^\mu}_5$ can cause to an effect which is very similar to the 
radiation 
effect that is calculated by Lorentz-Dirac equation.  The fact that the mass
is not conserved (the off-mass-shell case) is equivalent, in the case of
 radiation, 
to loss of energy through the radiation process. The particular example
that we treat here
is that of a charged particle in the presence of an uniform magnetic field in 
$z$ direction ($ {\bf B} = (0,0,B_0)$).
\par As for  the radiation reaction term of the Lorentz-Dirac 
equation, we choose the fifth field term to be 
\footnote{*}{It appears that for 
the usual form of the radiation reaction, in an example with the field
 magnitudes
that we shall choose, the ${d {\ddot x}^\mu \over d\tau}$ term seems to
 be negligible, and the ${\ddot x}^\mu
{\ddot x}_\mu$ may be approximated by a constant number; one is left with 
the $\dot x^\mu$ term.  We choose the fifth field term to have a similar structure.  This choice is appropriate due to the close relation of these  
 the radiation reaction terms of the usual theory [22].}
$${f^\mu}_5 = (C_1 \dot t, C_2 \dot x, C_2 \dot y, 0),
\eqno(A.1)$$
where the dot indicates derivative with respect to $\tau$.
The Lorentz force $(2.26)$ can be written as a set of differential equations,
$$M {{d^2t}\over{d\tau^2}} = {e\over c}C_1 {{dt}\over{d\tau}}\eqno(A.2) $$
$$M {{d^2x}\over{d\tau^2}} = {{eB_0}\over c}{{dy}\over{d\tau}}+
{e\over c}C_2 {{dx}\over{d\tau}} \eqno(A.3) $$
$$M{{d^2y}\over{d\tau^2}} = -{{eB_0}\over c}{{dx}\over{d\tau}}+
{e\over c}C_2 {{dy}\over{d\tau}}. \eqno(A.4)$$
The solution of Eq. $(A.2)$ is
$$\dot t =  \dot t_0 e^{\alpha_1\tau},\eqno(A.5)$$
where $\alpha_1={{eC_1}\over{Mc}}$. Using the complex coordinate 
$u=\dot x+i\dot y$, Eqs. $(A.3)$ and $(A.4)$ can be written as 
$${{du}\over{d\tau}}=-i\Omega u+\alpha_2 u, \eqno(A.6)$$
where $\alpha_2={{eC_2}\over{Mc}}$ and $\Omega={{eB_0}\over{Mc}}$ (the Larmor
frequency). The solution is
$$u = u_0 \exp^{\alpha_2\tau} e^{-i\Omega\tau}. \eqno(A.7)$$
Using $u(\tau)$ one finds that,
$$\eqalign{\dot x &= e^{\alpha_2\tau}(\dot x_0 \cos(\Omega\tau)+
\dot y_0 \sin(\Omega\tau)) \cr
\dot y &= e^{\alpha_2\tau}(-\dot x_0 \sin(\Omega\tau)+
\dot y_0 \cos(\Omega\tau)). \cr} \eqno(A.8)$$
As expected, the radiation effect is determined by the constants
$\alpha_1$ and $\alpha_2$.
\par It is possible to calculate the actual velocities by dividing Eqs. 
$(A.8)$ by $\dot t$; this results in
$$\eqalign{{{dx}\over{dt}} &= e^{-\alpha\tau}\left(\left({{dx}
\over{dt}}\right)_0 
\cos(\Omega\tau)+\left({{dy}\over{dt}}\right)_0 \sin(\Omega\tau)\right)\cr
{{dy}\over{dt}} &= e^{-\alpha\tau}\left(-\left({{dx}\over{dt}}\right)_0 
\sin(\Omega\tau)+\left({{dy}\over{dt}}\right)_0 \cos(\Omega\tau)\right),
\cr} \eqno(A.9)$$
where $\alpha=\alpha_1-\alpha_2$. Notice that when $\alpha_1=\alpha_2$, there
is apparent radiation (decrease of amplitude) as a function of
 $\tau$  but not as a function of $t$ ; in terms of $t$ (which is redshifted) 
the particle appears to be circling forever on the same circle.  This 
remarkable illustration is somewhat analogous to the phenomenon in which there
 is an infinite time required for a particle to arrive at the Schwarzschild
 radius in the Schwarzschild coordinate $t$, but a finite interval in the
 proper time of the particle.
\par The magnitude of the ($t$-) velocity of the particle is 
$$ v=v_0 e^{-\alpha\tau}. \eqno(A.10)$$
When $\alpha={1\over{\tau_0}}$, where $\tau_0={1\over{\gamma_0\Omega^2}}$
($\gamma_0$ is the radiation constant of the Lorentz-Dirac equation),
Eq. $(A.10)$ is exactly the solution which was obtained using the 
Lorentz-Dirac equation [14][15]. This result is consistent with the 
approximations we have made in constructing the example.
\vfill
\break
\noindent
{\it {\bf References}}
\frenchspacing
\smallskip 
 
\item{[1]} G.M. Zaslavskii, M.Yu. Zakharov, R.Z. Sagdeev,
D.A. Usikov, and A.A. Chernikov, Zh. Eksp. Teor. Fiz {\bf 91}, 500 (1986)
[Sov. Phys. JEPT {\bf 64}, 294 (1986)].
 
\item{[2]} V.I. Arnold'd, Dokl. Akad. Nauk. SSSR {\bf 159}, 9 (1964).
 
\item{[3]} D.W. Longcope and R.N. Sudan, Phys. Rev. Lett. {\bf 59},
1500 (1987); See also, A.J. Lichtenberg and M.A. Lieberman, {\it Regular and
Chaotic Dynamics} 2nd ed., (Springer-Verlag, New York, 1992).
 
\item{[4]} H. Karimabadi and V. Angelopoulos, Phys. Rev. Lett.
{\bf 62}, 2342 (1989).
 
\item{[5]} We thank T. Goldman for a discussion of this point.

\item{[6]} L.D. Landau and E.M. Lifshitz,{\it The Classical Theory of
Fields} 4th ed., (Pergamon Pr., Oxford, 1975). 
 
\item{[7]}F. Rohrlich, {\it Classical Charged Particles}, Addison
Wesley, Reading, (1965). 
 
\item{[8]} J.T. Mendon\c ca and L. Oliveira e Silva, Phys. Rev E {\bf
55}, 1217 (1997).
 
\item{[9]}  E.C.G. Stueckelberg, Helv. Phys. Acta {\bf 14}, 322
(1941); {\bf 14}, 588 (1941).

\item{[10]} L.P. Horwitz and C. Piron, Helv. Phys. Acta {\bf 46}, 316
(1973).

\item{[11]} B. Mashoon,Proc. VII Brazilian School of
Cosmology and Gravitation, Editions Fronti\'eres (1944);see also,
  Phys. Lett. A {\bf 145}, 147 (1990) and Phys. Rev.A{\bf 47}, 4498 (1993).
 We thank J. Beckenstein for bringing these references to our attention.
 
\item{[12]} M. Abraham, {\it Theorie der Elektrizit\"at}, vol. II,
Springer, Leipzig (1905).  See ref.[7] for a discussion of the origin
of these terms.

\item{[13]} P.A.M. Dirac, Proc. Roy. Soc. London Ser. A, {\bf 167},
148(1938).

\item{[14]}  A.A. Sokolov and I.M. Ternov, {\it Radiation from
Relativistic Electrons}, Amer. Inst. of
Phys. Translation Series, New York (1986).

\item{[15]} Y. Ashkenazy and L.P. Horwitz, Discrete Dyn. in Nature and 
Soc., this volume.
 
\item{[16]} A. Einstein, Phys. Z. {\bf 12}, 509 (1911).  See also
W. Pauli, {\it Theory of Relativity}, Dover, N.Y. (1981).

\item{[17]} L.P. Horwitz, R.I. Arshansky and A. Elitzur, Found. 
Phys. {\bf 18}, 1159 (1988).

\item{[18]} For example, S. Weinberg, {\it Gravitation and Cosmology:
Principles and Applications of the General Theory of Relativity},
Wiley, N.Y. (1972).

\item{[19]} R.I. Arshansky and L.P. Horwitz, Jour. Math. Phys. {\bf
30}, 66, 380, (1989).

\item{[20]} C. Piron, personal communication.

\item{[21]}  D. Saad, L.P. Horwitz and R.I. Arshansky, Found. of
Phys. {\bf 19}, 1125 (1989); M.C. Land, N. Shnerb and L.P. Horwitz,
Jour. Math. Phys. {\bf 36}, 3263 (1995); N. Shnerb and L.P. Horwitz,
Phys. Rev A{\bf 48}, 4068 (1993).

\item{[22]} O. Oron and L.P. Horwitz, in preparation.
 
\break

\noindent
Figure Caption
\bigskip
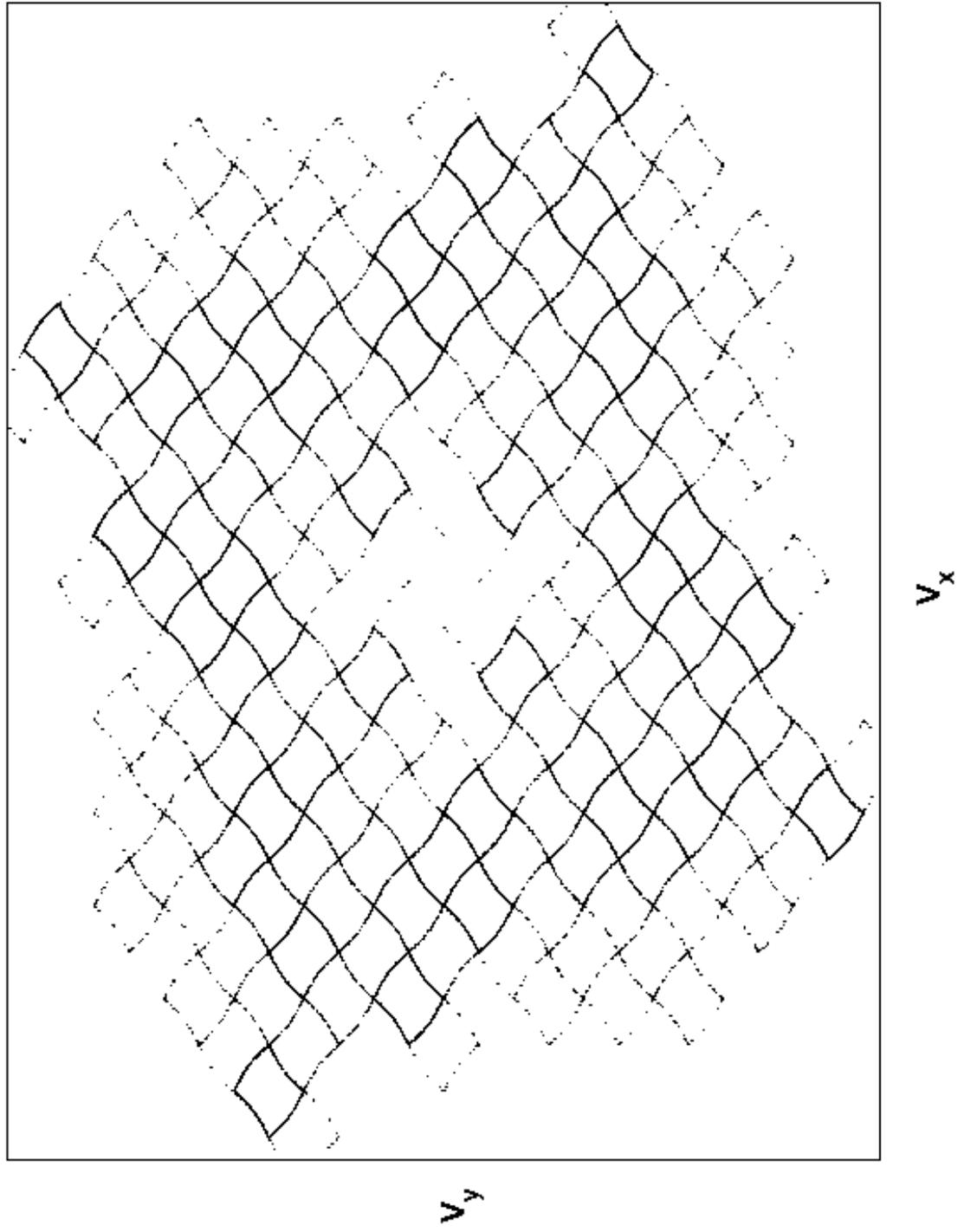
fig. 1. A typical relativstic stochastic web.
\vfill

\end
\bye